# Quantum cryptography - an information theoretic security

*Muhammad Nadeem*
*Department of Basic Sciences,*
*School of Electrical Engineering and Computer Science*
*National University of Sciences and Technology (NUST)*
*H-12 Islamabad, Pakistan*
*muhammad.nadeem@seecs.edu.pk*

Methods of quantum mechanics promise information-theoretic security for various protocols in cryptography. However, impossibility of some cryptographic applications such as standard bit commitment, oblivious transfer, multiparty secure computations and ideal coin tossing in quantum regime leaves an obvious question on the completeness of quantum cryptography. Instead of using wide range of rules and techniques for a variety of cryptographic applications, we demonstrate here a unified structure for quantum cryptography based on quantum non-local correlations. The unified framework achieves same goals in information-theoretic way as classical cryptography does with computational hardness. To cover the broad range of cryptographic applications, we show that the framework (i) assures secrecy by providing encryption completely unintelligible to eavesdroppers, (ii) guarantees that input from distant parties is concealed unless they are willing to reveal, (iii) assures binding, (iv) allows splitting information between several parties securely and more generally, (v) evades both quantum and classical attacks from internal as well as external eavesdropping.

**Introduction**
Quantum information theory[1] encodes information over probabilistic microscopic systems, called qubits, which are represented by unit vectors in Hilbert space and are processed through unitary operators. As a result, rules of quantum mechanics such as interference, uncertainty principle, quantum no-cloning[2], and EPR type correlations[3-5] provide novel features in information processing such as quantum teleportation[6], quantum computation[7,8], quantum communication[9,10], and quantum cryptography[11-13].

Among these advancements in information processing, quantum cryptography is the most promising and practically existed application of quantum mechanics. The most celebrated application of quantum cryptography, namely QKD[11,12], is the outcome of quantum uncertainty principle and quantum non-local correlations. QKD is the very basic ingredient for communication between distant parties with information theoretic security[14]. Besides cryptography based on QKD[11,12,14], a number of other information-theoretic quantum cryptographic applications such as optimal quantum encryption techniques[15], quantum secret sharing[16] (QSS), quantum finger printing[17], quantum digital signatures[18,19] (QDS), and asymmetric quantum cryptography in general have also been proposed.

However, a large class of cryptographic applications such as standard bit commitment (BC), oblivious transfer (OT, proposed by Rabin), two-sided two-party secure computation (TPSC), and asynchronous ideal coin tossing (CT) are still considered to be impossible in quantum cryptography[20-24]. Such no-go theorems for important tasks in quantum regime with information theoretic security leave an obvious question on the completeness of quantum cryptography.

Here we address the following very interesting questions: Instead of using wide range of techniques in quantum cryptography, is it possible to develop a unified quantum framework that would implement all tasks with information-theoretic security? A related question would be to ask: can such a general framework achieve same goals in information-theoretic way as classical cryptography does with computational hardness? If yes, then the framework needs to incorporate following security requirements: (i) it must provide perfect encryption completely unintelligible to eavesdroppers (secrecy). (ii) Input from distant parties must be concealed (BC, OT, MPSC, CT). (iii) It must assure binding (BC, QDS, MPSC, and CT). (iv) It must split information between several parties securely (QSS and QDS). More importantly, (v) it must evade both quantum and classical attacks from internal as well as external eavesdropping.

We demonstrate here that answers to the above questions are positive and propose a unified quantum framework, expressible in a single mathematical equation. The proposed framework is based on quantum non-local correlations in the form of teleportation[6] and entanglement swapping[6,25] and splits information between several parties where each of these parties extracts piece of secret that is non-locally correlated with information other parties have. We show that these non-locally correlated pieces of information can be used for implementing all pioneering cryptographic tasks such as standard bit commitment with arbitrarily long commitment time[26], quantum secret sharing[27], quantum digital signature scheme[28], oblivious transfer, two sided two-party secure computation, multiparty-party secure computation, and hence ideal coin tossing.

In a related work, we showed that theory of relativity adds more powers and extends scope of this unified framework. For example, relativistic version of proposed framework allows secure positioning[29-31], multiparty BC where both parties commit and reveal simultaneously[32,33], QSDC purely based on positioning, secure and authenticated communication based on bilateral secure positioning, information-theoretic security for *two-sided* cryptographic applications[34] (TPSC, CT), and relativistic OT where both data and position remains oblivious[34].

**Entanglement**

One of the most important resources of quantum information is entanglement. Detailed discussion about entanglement, its relation with foundations of quantum mechanics, and applications in quantum cryptography can be found in a recent review[35]. Let's denote Bell states as $\Psi^\mu = (\Psi^0, \Psi^i)$ with $\mu$ running over 0,1,2,3 while $i$=1,2,3. To keep the notations simple, we are not using Dirac's bta-ket notations for Bell states here. However, we keep using Dirac's notation for single qubit states.

$$\Psi^0 = \frac{|00\rangle + |11\rangle}{\sqrt{2}} \qquad (1.1)$$

$$\Psi^1 = \frac{|01\rangle + |10\rangle}{\sqrt{2}} \qquad (1.2)$$

$$\Psi^2 = \frac{|00\rangle - |11\rangle}{\sqrt{2}} \qquad (1.3)$$

$$\Psi^3 = \frac{|01\rangle - |10\rangle}{\sqrt{2}} \qquad (1.4)$$

Each of these four Bell states is related to other three by the unilateral application of one of the three Pauli matrices $\sigma_i$. For example, $\sigma_0 \otimes \sigma_i \Psi^0 = \Psi^i$ or we can write $\Omega_i \Psi^0 = \Psi^i$ where $\Omega_\rho = (\Omega_0, \Omega_i)$ are $2^2 \times 2^2$ unitary matrices with $Tr(\Omega_0) = 4$ and $Tr(\Omega_i) = 0$ and can be written in outer product form of $2 \times 2$ Pauli matrices $\sigma_\rho = (\sigma_0, \sigma_i)$ as $\Omega_\rho = \sigma_0 \otimes \sigma_\rho$.

$$\sigma_0 = I = \begin{pmatrix} 1 & 0 \\ 0 & 1 \end{pmatrix}, \sigma_1 = \sigma_x = \begin{pmatrix} 0 & 1 \\ 1 & 0 \end{pmatrix}, \sigma_2 = \sigma_z = \begin{pmatrix} 1 & 0 \\ 0 & -1 \end{pmatrix}, \sigma_3 = i\sigma_y = \sigma_z \sigma_x = \begin{pmatrix} 0 & 1 \\ -1 & 0 \end{pmatrix}$$

$$\Omega_0 = \begin{pmatrix} 1 & 0 & 0 & 0 \\ 0 & 1 & 0 & 0 \\ 0 & 0 & 1 & 0 \\ 0 & 0 & 0 & 1 \end{pmatrix}, \Omega_1 = \begin{pmatrix} 0 & 1 & 0 & 0 \\ 1 & 0 & 0 & 0 \\ 0 & 0 & 0 & 1 \\ 0 & 0 & 1 & 0 \end{pmatrix}, \Omega_2 = \begin{pmatrix} 1 & 0 & 0 & 0 \\ 0 & -1 & 0 & 0 \\ 0 & 0 & 1 & 0 \\ 0 & 0 & 0 & -1 \end{pmatrix}, \Omega_4 = \begin{pmatrix} 0 & 1 & 0 & 0 \\ -1 & 0 & 0 & 0 \\ 0 & 0 & 0 & 1 \\ 0 & 0 & -1 & 0 \end{pmatrix}$$

If we define inner product of $\Omega_\rho$ and $\Omega_{\rho'}$ as $Tr(\Omega_\rho \Omega_{\rho'}^T)$, then the set $\Omega_\rho = (\Omega_0, \Omega_i)$ forms orthonormal basis for four dimensional inner product space.

$$Tr(\Omega_\rho \Omega_{\rho'}^T) = 4\delta_{\rho\rho'} \tag{2}$$

Or

$$\sum_{\rho=0}^{3} \Omega_\rho \Omega_\rho^T = 4I \tag{3}$$

Moreover, one-to-one mapping of sets $\Omega_\rho = (\Omega_0, \Omega_i)$ and $\Psi^\mu = (\Psi^0, \Psi^i)$ leads to completely mixed outcome, that is,

$$\Omega_\rho \Psi^\mu = \Psi^0 \tag{4}$$

for $\rho = \mu$. For example, $\Omega_0 \Psi^0 = \Psi^0$, $\Omega_1 \Psi^1 = \Psi^0$ and so on.

**Quantum cryptography**

We demonstrate here a unified framework for quantum cryptography, expressible in a single mathematical equation, and show that it describes all pioneering cryptographic tasks such as standard bit commitment with arbitrarily long commitment time, quantum secret sharing, quantum digital signature scheme, oblivious transfer, deterministic multiparty-party secure computations and hence coin tossing. The rest of the cryptographic tasks can be built on any one of these fundamental procedures.

For generality, let's suppose there are three stations A, B and C who share a quantum system $H_i = |\psi\rangle \otimes \Psi^\mu \otimes \Psi^\nu$. Here $|\psi\rangle$ is a single qubit state kept at station A while $\Psi^\mu$ and $\Psi^\nu$ are maximally entangled Bell states shared between stations A and C and B and C respectively. By using Bell basis (1), initially shared quantum system $H_i$ can be written as

$$|\psi\rangle \otimes \Psi^\mu \otimes \Psi^\nu = \frac{1}{4} \left(\Omega_\tau \Omega_\rho \Psi^\mu\right)_a \otimes \left(\sigma^\tau |\psi\rangle\right)_b \otimes \left(\Omega^\rho \Psi^\nu\right)_c \tag{5}$$

where $\Psi''^\mu = \Omega_\tau \Omega_\rho \Psi^\mu$, $|\psi'\rangle = \sigma^\tau |\psi\rangle$, and $\Psi'^\nu = \Omega_\rho \Psi^\nu$. Above equation (5) is the set of 16 equations where Einstein summation rule is applied on index $\rho$ and $\tau$.

Now if station C performs Bell state measurement (BSM)[36] on $\Psi''^\nu$, he gets two classical bits $cc' \in \{00,01,10,11\}$ while distant and uncorrelated stations A and B gets entangled through one of the EPR channel $\Psi'^\mu \in \{\Psi^0, \Psi^i\}$. Now station A can teleport quantum state $|\psi\rangle$ to station B over EPR channel $\Psi'^\mu$ by performing BSM on $\Psi''^\mu$. As a result, A gets classical 2-bits $aa' \in \{00,01,10,11\}$ while entangled half at station B becomes $|\psi'\rangle = \sigma^\tau |\psi\rangle$.

For each value of C's BSM result $cc'$, there comes a unique Bell state $\Psi'^\mu$ swapped between A and B and hence unique teleportation encoding of quantum state $|\psi\rangle$ corresponding to BSM result $aa'$ of A. In other words, the teleportation encoding $\sigma^\tau$ is non-locally correlated with classical 2-bit strings $aa'$ and $cc'$ as well as initially shared EPR system $\Psi^\mu \otimes \Psi^\nu$. Hence, the proposed framework allows distributing non-locally correlated information at all three stations A, B, and C.

For generality, we formulated the task between three labeled stations A, B and C. For any specific cryptographic application, these stations can be assigned to different parties as follows: For quantum BC, OT, TPSC, and CT, station A is kept by committer/client Alice while stations B and C are controlled by receiver/server Bob. For QSS and QDSS, and TPSC, stations A, B, and C are controlled by three different parties namely Alice, Bob and Charlie where Alice is the sender while Bob and Charlie are receivers.

For $k$-bit string (massage/commitment/input) $|\psi\rangle \in \{0,1\}^{\otimes k}$, the proposed formulism can be iterated $k$ times where each iteration sends/commits 1-bit from station A to B and C or provide deterministic 1-bit outcome of two-sided computations $f(\sigma_a, \sigma_b, \sigma_c; |\psi\rangle)$ calculated over input state $|\psi\rangle$. Here Pauli transformation $\sigma_i$ is input from station $i$. Let's suppose all stations agree on the following codes:

**Code-I:** Pauli transformations $\sigma_0, \sigma_1, \sigma_2$ or $\sigma_3$ corresponds to classical 2-bit strings 00,01,10 or 11 respectively. That is, by applying one of the Pauli operators $\sigma_{ms} \in \{\sigma_0, \sigma_1, \sigma_2, \sigma_3\}$ to the shared quantum system $|\psi\rangle \otimes \Psi^\mu \otimes \Psi^\nu$ means an input of 2-bit string $ms \in \{00,01,10,11\}$ to the system.

**Code-II:** The mapping $ms \in \{00,10\} \to \{\sigma_0, \sigma_2\}$ represents signature bit $s=0$ while message bit $m \in \{0,1\}$. Similarly, the mapping $ms \in \{01,11\} \to \{\sigma_1, \sigma_3\}$ represents signature bit $s=1$ while message bit $m \in \{0,1\}$.

In other words, for each signature bit, there are two arbitrary message bits corresponding to Pauli operator $\sigma_{ms} = \sigma_z^m \sigma_x^s$. Hence, if the protocol is iterated $k$ time, the input $\sigma_{ms}^{\otimes k}$ generates $k$-bit signature, $n = 2^k$ possible $k$-bit messages, and $k$-bit deterministic outcome of some function $f(\sigma_a, \sigma_b, \sigma_c; |\psi\rangle)$ calculated over input state $|\psi\rangle \in \{0,1\}^{\otimes k}$.

**Information-theoretic security**

Unconditionally secure encrypting of $n$ qubit state $|\psi\rangle$, quantum one-time pad[15], is possible if encryption algorithm uses at least $2n$ random classical bits and the set $\{p_\rho, U_\rho\}$ with $p_\rho = 1/2^{2n}$ and $\{U_\rho\}$ is the set of $2^{2n}$ unitary $2^n \times 2^n$ matrices that form an orthonormal basis. The quantum one-time pad is information-theoretically secure if following condition holds:

$$|\psi'\rangle\langle\psi'| = \sum_{\rho=0}^{2^{2n}-1} p_\rho U_\rho |\psi\rangle\langle\psi| U_\rho^\dagger = \frac{1}{2^n} I \qquad (6)$$

Or

$$\sum_{\rho=0}^{2^{2n}-1} p_\rho U_\rho U_\rho^\dagger = I \qquad (7)$$

Our task here is not the unconditional secrecy of $|\psi\rangle$ from external eavesdropping only. This can be achieved from teleportation straightforward. Considering the generality of proposed framework, we would demand (i) perfect encryption completely unintelligible to external eavesdroppers, (ii) concealing from legitimate users, (iii) binding from sender, (iv) information splitting between several parties securely, and (v) finally information theoretic security from both classical and Mayers and Lo-Chau (MLC) quantum attacks[20-24].

To see whether proposed framework (5) satisfies all these security requirements or not, let's divide equation (5) into two parts; first entanglement swapping by station C and then teleportation by station A over swapped EPR channel between A and B as follows:

$$|\psi\rangle \otimes \Psi^\mu \otimes \Psi^\nu \xrightarrow{ES} \frac{1}{2}|\psi\rangle \otimes (\Omega_\rho \Psi^\mu)_{ab} \otimes (\Omega^\rho \Psi^\nu)_c \xrightarrow{Tele} \frac{1}{4}(\Omega_\tau \Omega_\rho \Psi^\mu)_a \otimes (\sigma^\tau |\psi\rangle)_b \otimes (\Omega^\rho \Psi^\nu)_c \qquad (8)$$

As standard for quantum cryptography, we assume that all the classical channels between Alice, Bob, and Charlie can be passively monitored but cannot be actively tempered by eavesdroppers. However, all the quantum channels can be actively disturbed.

*Entanglement swapping:* Station C shares Bell states $\Psi^\mu$ and $\Psi^\nu$ with stations A and B respectively while stations A and B are uncorrelated initially. Procedure of entanglement swapping can be written in Bell basis as

$$\Psi^\mu \otimes \Psi^\nu = \frac{1}{2}(\Omega_\rho \Psi^\mu)_{ab} \otimes (\Omega^\rho \Psi^\nu)_c \qquad (9)$$

Here Einstein summation rule is applied on index $\rho$. That is, initially shared quantum system $\Psi^\mu \otimes \Psi^\nu$ is local unitary equivalent to $\Psi'^\mu \otimes \Psi'^\nu$ where $\Psi'^\mu = \Omega_\rho \Psi^\mu$ is Bell state possessed by distant stations A and B while Bell state $\Psi'^\nu = \Omega^\rho \Psi^\nu$ is kept at station C. By applying local Bell operator on $\Psi'^\nu$, C gets two classical bits $cc'$ while distant and uncorrelated stations A and B gets entangled through one of the EPR channel $\Psi'^\mu \in \{\Psi^0, \Psi^i\}$. Regardless of the initially shared Bell states $\Psi^\mu$ and $\Psi^\nu$, outcome of Bell operator on state $\Psi'^\nu$ at station C can be one of the four possible 2-bit string $cc' \in \{00, 01, 10, 11\}$ each with probability $p_\rho = 1/4$.

$$\sum_{\rho=0}^{3} p_\rho \Omega_\rho \Omega_\rho^T = I \qquad (10)$$

Hence, the initially shared quantum system $\Psi^\mu \otimes \Psi^\nu$ carries no information about Bell state $\Psi'^\mu$ swapped between distance stations A and B unless A and B do not communicate with each other or with C.

$$\sum_{\rho=0}^{3} p_\rho \Omega_\rho (\Psi'^\mu)(\Psi'^\mu)^T \Omega_\rho^T = \frac{1}{4} I \qquad (11)$$

*Teleportation:* Now suppose station A keeps single qubit state $|\psi\rangle$ and shares a Bell state $\Psi'^{\mu} = \Omega_{\rho}\Psi^{\mu}$ with a distant station B. Initially shared quantum system $|\psi\rangle \otimes \Psi'^{\mu}$ can be rewritten in Bell basis as

$$|\psi\rangle \otimes \Psi'^{\mu} = \frac{1}{2}\Omega_{\tau}\Psi'^{\mu} \otimes \sigma^{\tau}|\psi\rangle \quad (12)$$

Here Einstein summation rule is applied on index $\tau$. That is, initially shared quantum system $|\psi\rangle \otimes \Psi'^{\mu}$ is local unitary equivalent to $\Psi''^{\mu} \otimes |\psi'\rangle$ where $\Psi''^{\mu} = \Omega_{\tau}\Psi'^{\mu} = \Omega_{\tau}\Omega_{\rho}\Psi^{\mu}$ is the Bell state on site of A while $|\psi'\rangle = \sigma^{\tau}|\psi\rangle$ is quantum state on site of B. By applying local Bell operator on state $\Psi''^{\mu} = \Omega_{\tau}\Psi'^{\mu}$, A can teleport quantum state $|\psi\rangle$ to B. Regardless of the initially shared quantum system $|\psi\rangle \otimes \Psi'^{\mu}$, outcome of Bell operator on state $\Psi''^{\mu}$ at station A can be one of the four possible 2-bit string $aa' \in \{00,01,10,11\}$ each with probability $p_{\rho} = 1/4$.

$$\sum_{\tau=0}^{3} p_{\tau}\Omega_{\tau}\Omega_{\tau}^{T} = I \quad (13)$$

Hence, the shared quantum system or qubit state $|\psi'\rangle$ at station B carries no information about the quantum state $|\psi\rangle$ unless A announces her BSM result and B and C collaborate.

$$\sum_{\tau=0}^{3} p_{\tau}\sigma_{\tau}|\psi\rangle\langle\psi|\sigma_{\tau}^{T} = \frac{1}{2}I \quad (14)$$

Here swapped EPR channel $\Psi'^{\mu}$ between A and B is known only to station C and remains unknown to both stations A and B unless C communicates with A or B. However, teleportation encoding $\sigma^{\tau}$ remains unknown to all three stations A, B and C. Now if station A publically announces her BSM result $aa'$, then teleportation encoding $\sigma^{\tau}$ is known only to C but still remains unknown to both A and B. Hence encrypted state $|\psi'\rangle = \sigma^{\tau}|\psi\rangle$ at station B remains arbitrary to A while message state $|\psi\rangle$ remains unknown to B even though he knows BSM result of A. However, if B and C collaborate, they can extract $|\psi\rangle$ from $|\psi'\rangle$ while A and C can know cipher state $|\psi'\rangle$ by mutual cooperation.

In short, if any station announces his/her share, other two stations can verify or aborts on the basis of their non-locally correlated information. The proposed framework fulfills all the security requirements (i)-(v) as follows:

*Secrecy from external eavesdropping:* The message state $|\psi\rangle$ is perfectly encrypted and fulfils security requirements of quantum one time pad (6,7). Hence any external eavesdropper E cannot excess state $|\psi\rangle$.

*Hiding from internal users:* The message state $|\psi\rangle$ remains hidden from both B and C unless A reveals her BSM result $aa'$. Even after getting $aa'$, station B remains ignorant about teleportation encoding $\sigma^{\tau}$ and hence message state $|\psi\rangle$.

*Binding:* Since station A does not know the ciphertext $\sigma^{\tau}|\psi\rangle$ stored at station B, he cannot synchronize her message $|\psi\rangle$ and BSM result $aa'$ with unknown ciphertext $\sigma^{\tau}|\psi\rangle$. Hence proposed framework provides information-theoretic security against repudiation by station A. Moreover, even if B knows the message when A reveals to him (QDSS), he cannot forge the message before sending to station C. In general, proposed framework provides binding to all stations where anyone applies Pauli transformation; he/she cannot repudiate or forge later.

***Information splitting:*** Can B or C extract $|\psi\rangle$ from $\sigma^\tau|\psi\rangle$ alone unless station A reveals? No. Ciphertext $\sigma^\tau|\psi\rangle$ is the outcome of two-fold encryption on message state $|\psi\rangle$: the unitary encryption operator $\sigma^\tau$ is non-locally correlated with BSM results $aa'$ and $cc'$ of stations A and C respectively. So only having ciphertext $\sigma^\tau|\psi\rangle$ or BSM result $cc'$ is not sufficient for finding message $|\psi\rangle$. However, if station A announces her BSM result $aa'$, B and C can collaborate and find message $|\psi\rangle$.

***MLC attacks:*** Can A, B or C exploit MLC attacks? That is, can any station delay his/her input and try to influence measurement outcome by getting information about other stations' input? No. Repetitive measurements and stored pieces of non-locally correlated classical information at various stations do not allow MLC attacks. The most obvious possibilities of exploiting MLC attacks can come from station C. If he does not perform BSM on $\Psi''^{iv}$, message state $|\psi\rangle$ will be teleported to him. To overcome this cheating strategy from C, station A can send her BSM result $aa'$ to station B only. Hence without $aa'$, C cannot extract $|\psi\rangle$ from cipher state $\sigma'^\tau|\psi\rangle$.

**Discussion:**
The proposed unified framework can be modified in a number of ways according to available quantum technologies: (i) instead of having pre-shared entanglement, each station can prepare and send entnagled qubit to the other station. This allows the host station to keep exact identity of epr pair secret and then built quantum cryptography with following code: The Bell states $\Psi^0, \Psi^1, \Psi^2, \Psi^3$ represent classical 2-bit string 00,01,10,11 respectively. That is, by preparing and sharing one of the Bell states $\Psi \in \{\Psi^0, \Psi^1, \Psi^2, \Psi^3\}$ means party is sending/committing corresponding 2-bit string $ms \in \{00,01,10,11\}$.

The same framework can be utilized for Asymmetric quantum cryptography by sharing a global public key $K_{\pi/2,\varphi_i} = \{(\pi/2, \varphi_i) : i = 1,2,3,...k\}$ and utilizing the quantum-classical correspondence between classical message $\psi = \psi_1\psi_2....\psi_k$ where $\psi_i \in \{0,1\}$ and quantum state $|\psi\rangle = |\psi_1\rangle|\psi_2\rangle....|\psi_k\rangle$. It provides an additional layer of security through completely mixed effects of Pauli operators on quantum state $K_{\pi/2,\varphi}|\psi\rangle$ and classical counterpart $\psi$. Moreover, it reduces the communication complexity; in most case, same outcomes can be achieved with less communication rounds.

Finally, theory of relativity adds more powers and extends scope of this unified framework. For example, relativistic version of proposed framework allows secure positioning, multiparty generalization of BC where both parties commit and reveal simultaneously, secure and authenticated direct communication based on bilateral secure positioning, information-theoretic security for *two-sided* cryptographic applications, and relativistic oblivious transfer where both data and position remains oblivious.

# Supplementary Material

## Applications in Cryptography

The proposed unified framework fulfills security requirements (i)-(v) and hence covers broad range of cryptographic primitives defined below:

**Bit commitment:** A bit commitment is a task between two mistrustful parties, a committer and a receiver. In general, committer commits to a specific bit by giving some information to the receiver and then unveils his/her commitment at some later time. Standard bit commitment is said to be information-theoretically secure if it fulfils following three security requirements: (i) Hiding: receiver should not be able to extract the committed bit value during the scheme. (ii) Binding: when committer reveals, it must be possible for receiver to know the genuine bit value with absolute guarantee while committer should not be able to reveal a bit different from the committed one. (iii) Indefinite commitment time: the scheme should sustain information-theoretic security for arbitrarily long time after commitment made by committer.

**Oblivious transfer:** OT was originally defined by Rabin where sender Bob sends a 1-bit message to the receiver Alice who can only receive the message with probability no more than half. The security of the protocol relies on the fact that Alice can find out whether or not he got the 1-bit message from Bob after the completion of protocol but Bob remains oblivious about it. In a related notion, 1-out-of-2 OT, Bob sends two 1-bit messages to Alice who can only receive one of them and remains ignorant about the other while Bob remains entirely oblivious to which of the two messages Alice received[1,2]. It was shown later by Crépeau that both of these notions of OT are equivalent[3].

**Asynchronous Ideal coin tossing:** Coin tossing[4] is another fundamental primitive function in communication that allows distant mistrustful parties Alice and Bob to agree on a random data. Coin tossing is said to be ideal if it follows: (i). It results in three possible outcomes $f : f_0 = 0$, $f_1 = 1$ or $f$ = invalid. (ii). Outcome $f_0$ and $f_1$ occurs with equal probability $P_{f_0} = P_{f_1} = 1/2$ and both parties have equal cheating probabilities, $P_a = P_b$, which means that the coin tossing is fair. (iii). If both parties are honest, the outcome $f$ = invalid never occurs; $P_{f=invalid} = 0$. (iv). If any one of the parties is dishonest, the outcome invalid occurs with probability $P_{f=invalid} = 1$.

**Two sided Two-party secure computations:** Two-sided TPSC enables two distant parties Alice and Bob to compute a function $f(a,b)$ where $a$ and $b$ are inputs from Alice and Bob respectively[5,6]. The protocol is said to be secure if it fulfils following security requirements: (i) both Alice and Bob learn output of $f(a,b)$ deterministically. (ii) Alice learns nothing about Bob's input $b$ except what logically comes from $a$ and $f(a,b)$. (iii) Bob learns nothing about Alice's input $a$ except what logically comes from $b$ and $f(a,b)$.

**Quantum secret sharing:** Secret sharing[7,8] is a cryptographic procedure where a secret (or decryption function of secret) is divided among multiple parties such that each party gets a part of secret (or decryption function of secret) but cannot extract the secret on its own. However, the secret can only be reconstructed (extracted) when all the parties collaborate together and share their part. Such a secret sharing scheme is called (*n*, *n*) threshold scheme. In general, the original secret can be reconstructed if and only if at least $k \leq n$ parties collaborate called a (*k*, *n*) secret sharing threshold scheme.

**Quantum digital signature:** A standard digital signature scheme[9,11] must fulfill at least following three security requirements: (a) the signature must be a pattern depending upon the message to be signed. (b) The signature must be built on some information unique to the signatory. (c) The scheme must allow all the recipients to store a copy of signature for verification at later stage.

All these pioneering cryptographic primitives have also been discussed in quantum cryptography where QSS[12-14] promises information splitting with information-theoretic security. However, standard BC[15-17], OT[18], asynchronous ideal CT[17], and two-sided TPSC[18,19] in general are considered to be impossible even in quantum regime. Methods of quantum mechanics, quantum one-way functions[20,21], guarantee information-theoretic security for quantum digital signature[21,22], but on the cost of quantum memory.

We demonstrate all these primitives with information-theoretic security and practically feasibility with current quantum technology; without requiring quantum memory or definite EPR pairs which have very low probability of generation. In all the cryptographic application outlines below, the first three steps of unified framework (5) will be common:

1. Station C performs BSM on $\Psi'^\nu$ and stores classical result $cc'$.

2. Station A teleports state $|\psi\rangle$ to B by performing BSM on $\Psi''^\mu$. As a result, station A gets classical 2-bit string $aa'$ while entangled half at station B becomes $|\psi'\rangle = \sigma^\tau |\psi\rangle$.

3. Station B measures $|\psi'\rangle$ and stores classical result $\psi' \in \{0,1\}$.

**Two-party cryptography**

Suppose stations A is kept by committer Alice while those of B and C are controlled by receiver Bob.

**Bit commitment:** For devising quantum bit commitment scheme, we assume that Bell state $\Psi^\mu$ is publically known, Bell state $\Psi^\nu$ is the secret of Bob while $|\psi\rangle$ where $\psi \in \{0,1\}$ is the secret bit committed by Alice.

4. Alice also sends state $|\psi''\rangle = \sigma_z^a \sigma_x^{a'} |\psi\rangle$ to Bob.

5. Bob measures $|\psi''\rangle$ and stores $\psi'' \in \{0,1\}$.

6. Alice reveals her commitment $\psi \in \{0,1\}$ to Bob.

*Verification:* Bob finds the 2-bit string $aa'$ from $|\psi''\rangle = \sigma_z^a \sigma_x^{a'} |\psi\rangle$. He verifies whether the commitment is genuine or not as follows: He finds $\sigma^\tau$ from non-locally correlated classical information $aa'$ and $cc'$. Bob authenticates the commitment genuine if $|\psi'\rangle = \sigma^\tau |\psi\rangle$ otherwise detects cheating. This simple bit commitment scheme fulfills all three security requirements of standard bit commitment: (i) hiding, (ii) binding, and (iii) indefinite commitment time.

**Asynchronous Ideal coin tossing:** Suppose initially shared system $\Psi^\mu \otimes \Psi^\nu = \Psi^0 \otimes \Psi^0$ is publically known. For an asynchronous ideal coin tossing, Alice and Bob want to compute the function $f(\sigma_c, \sigma^\tau; |\psi\rangle)$ where $|\psi\rangle \in \{0,1\}$ is the secret input by Alice and $\sigma_c = \sigma_z^c \sigma_x^{c'}$ is transformation applied by Bob on state $|\psi'\rangle = \sigma^\tau |\psi\rangle$.

4. Bob applies transformation $\sigma_c = \sigma_z^c \sigma_x^{c'}$ on $|\psi'\rangle = \sigma^\tau |\psi\rangle$ and announces the outcome of coin tossing $f(\sigma_c, \sigma^\tau; |\psi\rangle) = \sigma_c \sigma^\tau |\psi\rangle$.

*Verification:* Alice applies unitary transformations $\sigma_a = \sigma_z^a \sigma_x^{a'}$ on function $f$ and verifies the outcome valid if $\sigma_a f = |\psi\rangle$ otherwise aborts. Since Bob do not know $|\psi\rangle$ and $aa'$, hence, he cannot simulate alterations in $cc'$ with $|\psi'\rangle = \sigma^\tau |\psi\rangle$ and hence outcome of his choice.

It fulfills the security requirements of an asynchronous ideal quantum coin tossing with zero bias: (i) deterministic outcome occurs with equal probability of $f = 0$ or $f = 1$; $P_{f=0} = P_{f=1} = 1/2$. (ii) Cheating probabilities for Alice and Bob are zero; $P_a = P_b = 0$. (iii) Probability $P_{f=invalid} = 0$ if both parties are honest and $P_{f=invalid} = 1$ if any one of the parties tries cheating.

The basic idea behind setting $\Psi^\mu \otimes \Psi^\nu = \Psi^0 \otimes \Psi^0$ and using step 4 is equation (4); $\Omega_\rho \Psi^\mu = \Psi^0$ if $\rho = \mu$. In other words, whatever the BSM result Alice and Bob get at stations A and C, following identity holds

$$\sigma_z^a \sigma_x^{a'} \sigma_z^c \sigma_x^{c'} \sigma^\tau = I \qquad (1)$$

Hence, without knowing each other's BSM results, both Alice and Bob get deterministic outcome of coin tossing. This simple coin tossing scheme is both ideal and asynchronous since no classical communication is required at all and both parties apply unitary transformations asynchronously.

**Oblivious transfer:** Suppose initially shared system $\Psi^\mu \otimes \Psi^\nu = \Psi^0 \otimes \Psi^0$ is publically known. For oblivious transfer from Bob to Alice, Alice and Bob want to compute the function $f(\sigma_c, \sigma^\tau; |\psi\rangle)$ where $|\psi\rangle \in \{0,1\}$ is the secret input by Alice and $\sigma_c = \sigma_z^c \sigma_x^{c'}$ is transformation applied by Bob on state $|\psi'\rangle = \sigma^\tau |\psi\rangle$.

4. Bob applies transformation $\sigma_c = \sigma_z^c \sigma_x^{c'}$ on $|\psi'\rangle = \sigma^\tau |\psi\rangle$ and sends the outcome of function $f(\sigma_c, \sigma^\tau; |\psi\rangle) = \sigma_c \sigma^\tau |\psi\rangle$ to Alice.

*Verification:* Alice applies unitary transformations $\sigma_a = \sigma_z^a \sigma_x^{a'}$ on function $f$ and verifies the outcome valid if $\sigma_a f = |\psi\rangle$ otherwise challenge the process. Both the security requirement of OT are satisfied: After the completion of protocol, (i) Bob do not know the Alice bitinput $|\psi\rangle \in \{0,1\}$ and (ii) Alice do not know the Bob's messages $c \in \{0,1\}$ corresponding to his signature $c'$.

**Two sided Two-party secure computations:** Suppose Station A is kept by Alice while B and C are controlled by Bob respectively who wants to compute the function $f(\sigma_a, \sigma_b, \sigma^\tau; |\psi\rangle)$ where $|\psi\rangle$ ($|\psi\rangle = |0\rangle$ or $|\psi\rangle = |1\rangle$) is publically known and Pauli transformations $\sigma_a$ and $\sigma_b$ on quantum system $|\psi\rangle$ are inputs from Alice and Bob respectively. To allow both parties giving their inputs, we rewrite steps 2 and 3 and then proceed as follows:

2. Alice applies Pauli transformation $\sigma_a \in \{\sigma_z^{a_1} \sigma_x^{a_2}, \sigma_z^{1 \oplus a_1} \sigma_x^{a_2}\}$ on $|\psi\rangle$, teleports the state $\sigma_a |\psi\rangle$ to Bob over EPR channel $\Psi'^\mu$. Simultaneously she sends $|\psi''\rangle = \sigma_z^a \sigma_x^{a'} \sigma_a |\psi\rangle$ to Bob where $\sigma_a' = \sigma_z^a \sigma_x^{a'}$ is depending on her BSM result $aa'$.

3. Bob measures his entangled half $|\psi'\rangle = \sigma^\tau \sigma_a |\psi\rangle$ and state $|\psi''\rangle = \sigma_z^a \sigma_x^{a'} \sigma_a |\psi\rangle$ and stores classical results. He applies Pauli transformation $\sigma_b \in \{\sigma_z^{b_1} \sigma_x^{b_2}, \sigma_z^{1 \oplus b_1} \sigma_x^{b_2}\}$ corresponding to his input and $\sigma'_c = \sigma_z^c \sigma_x^{c'}$ on $|\psi'\rangle$ and sends $|\psi'''\rangle = \sigma'_c \sigma_b \sigma^\tau \sigma_a |\psi\rangle$ to Alice.

4. Alice applies $\sigma'_a = \sigma_z^a \sigma_x^{a'}$ on $|\psi'''\rangle$ and announces both his BSM result $aa'$ and the outcome of two-sided computation

$$f(\sigma_a, \sigma_b; |\psi\rangle) = \sigma'_a \sigma'_c \sigma_b \sigma^\tau \sigma_a |\psi\rangle \qquad (2)$$

It fulfills the security requirements of secure and deterministic two-sided TPSC: (i) Alice can find Bob's transformation $\sigma_b \to b_1 b_2$ but his input $b_1 \in \{0,1\}$ remains totally random to her. (ii) Similarly, Bob can find Alice's transformation $\sigma_a \to a_1 a_2$ but her input $a_1 \in \{0,1\}$ remains totally random to him. Finally, (iii) both Alice and Bob gets same unbiased outcome of function $f(\sigma_a, \sigma_b; |\psi\rangle)$ deterministically.

**Multi-party cryptography**

Suppose stations A, B, and C are controlled by three independent parties namely Alice, Bob, and Charlie respectively.

**Quantum secret sharing:** The proposed framework results in (2, 2) secure quantum secret sharing that allows sender Alice to encrypt her secret $|\psi\rangle$, which can be either classical $\psi \in \{0,1\}$ or purely quantum mechanical $|\psi\rangle = \alpha|0\rangle + \beta|1\rangle$, such that only a group of receivers (Bob and Charlie) can decrypt her secret from $\sigma^\tau |\psi\rangle$.

If Alice follows procedure (5), $\sigma^\tau |\psi\rangle$ in possession of receiver Bob gives perfect encryption of her secret $|\psi\rangle$ where decryption function $\sigma^\tau$ is non-locally correlated with both BSM results $cc'$ and $aa'$ of receiver Charlie and sender Alice respectively. As a result, secret $|\psi\rangle$ can be decoded from $\sigma^\tau |\psi\rangle$ if and only if Alice reveals $aa'$ while Bob and Charlie collaborate with each other.

While considering the possibilities of MLC quantum attacks by Charlie, sender Alice needs to assure authentication from receiver's ends. Before revealing her BSM result, she would require some sort of authentication tokens from both Bob and Charlie and assurance that state $|\psi\rangle$ has been teleported to Bob and not Charlie. In other case, Charlie can have both shares and hence can find secret alone.

**Quantum digital signature:** For devising quantum digital signature scheme based on quantum secret sharing, we assume that publically known Bell states $\Psi^\mu$ and $\Psi^\nu$ are pre-shared while $|\psi\rangle$ ($\psi \in \{0,1\}$) is the message to be signed by Alice. Here, non-locally correlated states $|\psi'\rangle = \sigma^\tau |\psi\rangle$ and $|\psi''\rangle = \sigma'_a |\psi\rangle$ where $\sigma'_a = \sigma_z^a \sigma_x^{a'}$, are considered to be signatures of Alice here.

4. Alice sends state $|\psi''\rangle = \sigma_z^a \sigma_x^{a'} |\psi\rangle$ to Charlie.

5. Charlie measures $|\psi''\rangle$ in $\{0,1\}$ basis and stores outcome.

6. Bob and Charlie share their secrets $|\psi'\rangle$ and $cc'$ with each other, secretly from Alice.

*Verification:* Alice sends her message $|\psi\rangle$ and BSM result $aa'$ to Bob. By using $aa'$ and $cc'$, Bob verifies whether $|\psi\rangle$ and $|\psi'\rangle$ are consistent with $\sigma^\tau$ or not. Bob authenticates the message if $|\psi'\rangle = \sigma^\tau|\psi\rangle$ otherwise detects repudiation. If Bob validates that message and signature are agreed, he forwards the message $|\psi\rangle$ to Charlie.

Now Charlie verifies whether forgery or repudiation has occurred or not as follows: He finds $aa'$ from $|\psi''\rangle = \sigma_z^a \sigma_x^{a'}|\psi\rangle$ and verifies whether $|\psi'\rangle = \sigma^\tau|\psi\rangle$ is consistent with $cc'$ or not. If teleportation encoding $\sigma^\tau$ is consistent with non-locally correlated shares $aa'$ and $cc'$, he accepts the message genuine.

For $k$-bit message $|\psi\rangle = |\psi_1\rangle|\psi_2\rangle....|\psi_k\rangle$, Bob and Charlie can exchange some parts of their secrets instated of exchanging whole shares. For example, Bob divides his secret $|\psi'\rangle = |\psi_1'\rangle|\psi_2'\rangle....|\psi_k'\rangle$ into two parts; $|\psi'\rangle_b$ and $|\psi'\rangle_c$ and sends $|\psi'\rangle_c$ to Charlie. Similarly, Charlie divides his secret $c = \{c_i c_i'; i=1,2,....k\}$ into $c_b$ and $c_c$ and sends $c_b$ to Bob. Here order of sequence $b$ and $c$ are agreed between Bob and Charlie but unknown to Alice. This prevents Alice to create conflict between Bob and Charlie.

**Multi-party secure computations:** Suppose system $H_i = |\psi\rangle \otimes \Psi^\mu \otimes \Psi^\nu$ is publically known while Alice, Bob, and Charlie want to compute the function $f(\sigma_a, \sigma_b, \sigma_c; |\psi\rangle)$. Here $\sigma_a$, $\sigma_b$ and $\sigma_c$ are unitary transformations on quantum system $|\psi\rangle \otimes \Psi^\mu \otimes \Psi^\nu$ applied by Alice, Bob and Charlie respectively. To allow all three parties giving their inputs, we rewrite steps 1-3 and then proceed as follows:

1. Charlie performs BSM on $\Psi'^\nu$ and stores his BSM result $cc' = c_1 c_2$.

2. Alice applies Pauli transformation $\sigma_a \in \{\sigma_z^{a_1}\sigma_x^{a_2}, \sigma_z^{1\oplus a_1}\sigma_x^{a_2}\}$ on $|\psi\rangle$, teleports the state $\sigma_a|\psi\rangle$ to Bob over EPR channel $\Psi'^\mu$ and announces her BSM result $aa' \in \{00,01,10,11\}$.

3. Bob measures his half $|\psi'\rangle = \sigma^\tau \sigma_a|\psi\rangle$, applies Pauli transformation $\sigma_b \in \{\sigma_z^{b_1}\sigma_x^{b_2}, \sigma_z^{1\oplus b_1}\sigma_x^{b_2}\}$ on $|\psi'\rangle$ and sends $|\psi''\rangle = \sigma_b \sigma^\tau \sigma_a|\psi\rangle$ to Charlie.

4. Charlie applies teleportation encoding $\sigma_c \in \{\sigma_z^{c_1}\sigma_x^{c_2}, \sigma_z^{1\oplus c_1}\sigma_x^{c_2}\}$ on $|\psi''\rangle$ and announces the outcome of two-sided computation

$$f(\sigma_a, \sigma_b, \sigma_c; |\psi\rangle) = \sigma_c \sigma_b \sigma^\tau \sigma_a |\psi\rangle \qquad (3)$$

*Verification:* All the parties announce their signatures $a_2$, $b_2$ and $c_2$ respectively. Now each of these parties can verify the outcome individually. Remember, announcements of signature does not reveal the input message; for each signature bit, there are two arbitrary message bits corresponding to Pauli operator $\sigma_{ms} = \sigma_z^m \sigma_x^s$.